\documentclass[a4paper, 12pt]{article}
\usepackage[a4paper, total={6.5in, 9.5in}]{geometry}
\usepackage{lineno}
\usepackage{dcolumn}
\usepackage{appendix}
\usepackage[utf8]{inputenc}
\usepackage{bm}
\usepackage[margin=1.5cm]{caption}
\usepackage{braket}
\usepackage{amsmath}
\usepackage{amsfonts}
\usepackage{mathtools}
\usepackage{graphicx}
\usepackage{cancel}
\usepackage{slashed}
\usepackage{wrapfig}
\usepackage{enumitem}
\usepackage{float}
\usepackage{hyperref}
\usepackage{amssymb}
\usepackage{enumitem}
\hypersetup{colorlinks=true, urlcolor=black, citecolor=blue, linkcolor=black}

\usepackage{graphicx} 
\usepackage{authblk}

\begin{document}

\title{\textbf{Entanglement Entropy for the Black 0-Brane}}
\date{}
\author[1]{Angshuman Choudhury \thanks{angshuman.choudhury@hertford.ox.ac.uk}}
\author[1]{Davide Laurenzano \thanks{davide.laurenzano@physics.ox.ac.uk}}
\affil[1]{ Rudolf Peierls Centre for Theoretical Physics, University of Oxford, Parks Road, Oxford OX1 3PU, United Kingdom}

\maketitle
\thispagestyle{empty}
\begin{abstract}
    We analyze the entanglement entropy between the Black 0-Brane solution to supergravity and its Hawking radiation. The Black 0-Brane admits a dual Gauge theory description in terms of the Matrix model for M-Theory, named BFSS theory, which is the theory of open strings on a collection of $N$ $D_0$-branes. Recent studies of the model have highlighted a mechanism of Black Hole evaporation for this system, based on the chaotic nature of the theory and the existence of flat directions. This paper further explores this idea, through the computation of the von Neumann entropy of Hawking radiation. In particular, we show that the expected Page curve is indeed reproduced, consistently with a complete recovery of information after the Black Hole has fully evaporated. A pivotal step in the computation is the definition of a Hilbert space which allows for a quantum mechanical description of partially evaporated Black Holes. We find that the entanglement entropy depends on the choice of a parameter, which can be interpreted as summarizing the geometric features of the Black Hole, such as the size of the resolved singularity and the size of the horizon. 
    \end{abstract}
    \maketitle
\newpage

\tableofcontents 
\section{Introduction}
\setcounter{page}{1}
Black Hole physics presents several unsolved, puzzling questions, and it is widely believed that answering those questions would shed light on the quantum nature of gravity. In particular, the problem of finding a unitary process describing the formation and evaporation of a Black Hole is pivotal for solving the Black Hole information paradox \cite{InfoParadox}, a longstanding problem in modern physics. This idea can be made concrete in the context of the Gauge/Gravity duality conjecture \cite{Maldacena_1999, witten1998anti}, in which one can study Black Hole formation/evaporation in the dual gauge theory. In this spirit, we focus in this note on the BFSS Matrix Theory. This is known to represent the low energy limit of string theory on a stack of $D_0$-branes. BFSS theory has been conjectured to describe the Discrete Light Cone Quantization (DLCQ) of M-Theory \cite{BFSS, susskind1997conjecture} and many consistency checks have been performed since the formulation of the duality. See e.g. \cite{Taylor_2001, Banks_1997, bigatti1997review,ydri2018review} for reviews. In this interpretation, M-theory objects, including low energy degrees of freedom of 11-dimensional supergravity, are mapped into bound states of $D_0$-branes, held together by open strings stretched between them. In particular, multi-particle states are represented by block diagonal matrix configurations, which are possible because the bosonic Matrix Theory potential has flat directions.

From a Gauge/Gravity duality viewpoint, the state in which all the branes and open strings form a single bound state, with eigenvalues clumped around the origin, has been conjectured \cite{Itzhaki_1998} to be dual to a Black 0-brane or Black Hole configuration in the 't Hooft large $N$ limit, where the matrix size $N$ goes to infinity and the 't Hooft coupling $\lambda=g^2_{YM}N$ is kept fixed. Deviations from the large $N$ regime represent stringy corrections to the Black Hole.  Within this interpretation, the instability due to the flat directions of the Matrix Theory potential is interpreted in terms of Hawking radiation, namely the emission of $D_0$-branes which escape from the bound state. Such a picture makes sense in light of the M-theory interpretation of BFSS theory, in which emitted branes are mapped into massless particles, consistently with the idea of a Black Hole emitting massless Hawking radiation.

Recently, it has been shown \cite{Berkowitz_2016, Hanada_2016} that the emission of $D_0$-branes in BFSS theory can be characterized by the classical (high temperature) limit of the Matrix Theory. The argument is based on the existence of flat directions in the potential and the fact that the dynamic of the theory is chaotic. The result is that Black Hole evaporation can be fully resolved in terms of the emission of branes in the Matrix Theory. The consequence is that the Black 0-brane should not feature any loss of information after complete evaporation. This paper aims to make this statement precise and show that it is indeed correct, by looking at the entanglement entropy between the Black Hole and its radiation. A key step will be to provide a quantum Hilbert space description of the Black Hole evaporation process, which will then allow us to compute the entanglement entropy. The construction of the Hilbert space will be similar to the one in \cite{deboer2024page}. The main difference with our work is the implementation of time evolution on the Hilbert space: while the authors of \cite{deboer2024page} consider the average over an ensemble of Hamiltonians to show that, in general, unitarity is not a necessary ingredient for the recovery of information after the Black Hole has evaporated, we consider unitary time evolution for a specific theory. 
The entanglement entropy in Matrix Theory was studied in \cite{Gautam2022} from the gauge theory point of view, where the extended Hilbert space can be factorized. In the present work, the factorization of the Hilbert space will be carried out in the gravitational theory. Our central result will be to show that the profile of the von Neumann entropy as a function of time follows a Page curve, implying that the information is fully recovered, after complete evaporation.

The paper is organized as follows. In section \ref{sec:BHeva}, we summarize the mechanism put forth in \cite{Berkowitz_2016} for Black Hole formation and evaporation. In section \ref{sec:QuantSystem}, we characterize the Hilbert space of the quantized Black Hole plus radiation system. This will enable us to compute the entanglement entropy. The computation is carried out in section \ref{sec:PageCurve}, in which the Page curve for the entropy is obtained. A more quantitative and model-dependent description of Hawking radiation is presented in section \ref{sec:Eigendensity}. This is based on the distribution of the largest eigenvalue of the radial coordinate matrix, as proposed in \cite{Berkowitz_2016}. Finally, we conclude with a brief summary of the results and outlook in section \ref{sec:Conclusion}.
\section{A Mechanism for BFSS Black Hole Evaporation}
\label{sec:BHeva}
As outlined in the introduction, a duality exists that maps the BFSS gauge theory into a theory of gravity. Within this framework, the dual description of a bound state of $N$ $D_0$-branes on the gauge theory side is a Black 0-brane or Black Hole, on the gravity side. As such, BFSS theory provides a framework to study a Black Hole configuration for which micro-states can be represented in terms of Matrix Theory quantities. 
BFSS theory is the dimensional reduction of (9+1)-dimensional Super Yang-Mills theory to 0+1-dimensions, and the Lagrangian of the theory reads
\begin{equation}
    \mathcal{L}=\frac{1}{2 g^2_{YM}}\operatorname{Tr}\Bigl[\bigl(D_t X^I\bigr)^2+\frac{1}{2}[X^I, X^J]^2+i \Bar{\psi}^{\alpha}D_t \psi_{\alpha}+\Bar{\psi}^{\alpha}\Gamma^I_{\alpha \beta}[X^I, \psi^{\beta}]\Bigr] \, .
    \label{eq:BFSS}
\end{equation}
Here $X^I, \, I=1,...,9$ are $N\times N$ bosonic hermitian matrices, $D_t=\partial_t-i[A_t, \cdot]$ is the covariant derivative, $A_t$ is the $U(N)$ gauge field, $\psi^{\alpha}, \, \alpha=1,...,16$ are $N \times N$ real fermionic matrices and $\Gamma^I_{ \alpha \beta}$ are the left hand part of gamma matrices in (9+1)-dimensions. When the matrices $X^I$ are close to diagonal, the entries on the diagonal can be thought of as describing the coordinates of the $N$ $D_0$-branes in spacetime, whereas off-diagonal terms represent interactions through open string stretched between branes. The bosonic potential, namely 
\begin{equation}
    V=\frac{1}{4 g^2_{YM}}\operatorname{Tr}\Bigl[[X^I, X^J]^2\Bigr] 
    \label{eq:BFSSpotential}
\end{equation}
has flat directions in which it vanishes. This allows for multi-particle states in the M-Theory picture which are represented by block diagonal matrices in the Gauge theory side. The presence of flat directions is also pivotal in the description of the evaporation mechanism, as presented below.

In the rest of the section, we summarise important features of the BFSS Black Hole, such as a mechanism for Black Hole formation and evaporation. The presentation will follow the arguments put forth in \cite{Berkowitz_2016}. 
\subsection{Black Hole Formation}
The formation of the BFSS Black Hole state, namely a single bound state of $N$ $D_0$-branes clumped around the origin and kept together by open strings, is made possible by the fact that it is entropically favoured compared to a collection of $N$ non-interacting branes. This is explained by looking at the bosonic part of the BFSS Lagrangian and, in particular, considering the matrices $X^I$, $I=1,...9$. When $D_0$-branes are non-interacting, the matrices $X^I$ are diagonal, thus sitting in the flat direction of the potential (\ref{eq:BFSSpotential}). Considering that those are valued in the adjoint of $SU(N)$, the number of degrees of freedom for such a non-interacting configuration is $O(N)$. On the other hand, if the $D_0$-branes are all interacting with one another and form a bound state, off-diagonal elements are non-zero and the eigenvalues (representing the position of branes) are clumped around the origin. This corresponds to the case in which the matrices are non-commuting, yielding a non-zero interaction potential. In this case, each matrix contributes $O(N^2)$ degrees of freedom. It follows that such a configuration is entropically favoured, being accounted for by a larger number of degrees of freedom.

Black Hole formation is, therefore, entropically favoured. Moreover, the matrix model is chaotic \cite{Berkowitz_2016} meaning that, for almost all initial conditions, the entire phase space will be explored by the system over time. Therefore, eventually, the system will explore the bound state described above, leading to Black Hole formation.
\subsection{Black Hole Evaporation}
The same chaotic nature of the theory causes the Black Hole to eventually evaporate. Since the system is ergodic, a single $D_0$-brane can explore configurations in which it is located far away from the bound state of the remaining $N-1$ branes. This causes the degrees of freedom describing the interaction of the $D_0$-brane with all other bounded $D_0$-branes to become heavy, as their mass is proportional to the length of the string stretched between the branes. When the mass becomes larger than the temperature, the classical treatment of the theory (\ref{eq:BFSS}) doesn't hold anymore and one loop effects become relevant. In particular, it has been shown \cite{BFSS,Kabat_1998}, that at large distances a cancellation occurs: attractive and repulsive forces compensate, heavy degrees of freedom decouple from the dynamics and can be integrated out to obtain an effective description of the system (the comparison between such an effective theory and graviton scattering in 11-dimensional supergravity has provided a perturbative check of the BFSS conjecture). Such a cancellation relies on contributions coming from both the fermionic and the bosonic part of the BFSS potential. On reaching a certain threshold, the interaction becomes very feeble and can be considered to be turned off. We then say that a flat direction of the Matrix Theory potential (\ref{eq:BFSSpotential}) opens up. We consider the $D_0$-brane to depart from the Black Hole when this happens, as the off-diagonal terms in the corresponding row and column of the $X^I$ matrices are vanishing, namely
\begin{equation}
X_{\mathrm{BH}}=\left(\begin{array}{cc}
X^{\prime} & x_I \\
x_I^{t *} & x_{D_0}
\end{array}\right)\xrightarrow[\text{decoupling}]{\text{on}}X_{\mathrm{BH}+D_ 0}=\left(\begin{array}{cc}
X^{\prime} & 0 \\
0 & x_{D_0}
\end{array}\right) \, ,
\end{equation}
where $X'$ is the $(N-1)\times(N-1)$ matrix describing the remaining $N-1$ $D_0$-brane bound state/Black Hole. Hence, we can consider the decoupled $D_0$-brane to be emitted by Hawking radiation from the evaporating Black Hole. 

The loss of the radiated $D_0$-brane brings about a loss of entropy of the Black Hole of the order $O(N)$, due to the decreasing of degrees of freedom, which implies that this process is increasingly improbable as $N$ increases. This has been verified by \cite{Berkowitz_2016}, as the authors argue that the emission rate of a $D_0$-brane from a $N$ $D_0$-brane Black Hole is $\approx e^{-N}$, implying that the emission of a $D_0$-brane is suppressed at large $N$. It is also argued that the emission of individual branes is dominant compared to the emission of composite states of $k$ $D_0$-branes with $1<k<N$. As a consequence, evaporation occurs as a sequence of single-brane emissions.

In the next section, we will provide a quantized description of the process of Black Hole evaporation which is consistent with this thermodynamic treatment. This will be the first step towards a derivation of the Page Curve for the entanglement entropy of Hawking radiation.
\section{Quantized System Description}
\label{sec:QuantSystem}
The BFSS Black Hole, as described above, is characterized as a bound state of $D_0$-branes. In this scenario, Hawking radiation is completely resolved as a unitary process consisting of the emission of branes. As a consequence, the evaporation process for such a Black Hole should not feature any loss of information. The aim of the following sections will be to confirm this statement. To do so, we will provide a quantum Hilbert space description for the emission of $D_0$-branes as a radiative process and we will calculate the von Neumann entanglement entropy of Hawking radiation as a function of time. The expectation is that this should reproduce a Page curve, consistent with the recovery of information from Hawking radiation. We will show that this is indeed the case. As a first step, we need to define a suitable Hilbert space that allows for entangled states of Black Hole and radiation. This will enable us to construct the radiation density matrix and, hence, to obtain the entanglement entropy for the radiation.

The mechanism for Black Hole evaporation outlined in section \ref{sec:BHeva} can be given a Hilbert space description in a quantum mechanical setting. In particular, we can consider the Hilbert space proposed in \cite{deboer2024page}, which we summarize in the following. The total Hilbert space can be decomposed in sub-spaces spanned by eigenstates of the Hamiltonian with energies taking values in small micro-canonical windows $\left[E-\delta E,E\right]$ with $\delta E\ll E$, namely
\begin{equation}
    \mathcal{H}=\bigoplus_{E}\mathcal{H}_{E, \delta E} \, .
\end{equation}
The small smearing of the energy ensures that we do not run into any divergences. Moreover, conservation of energy implies that, once we fix the energy of the system, the dynamics will never leave the corresponding Hilbert subspace.  We can, therefore, focus on a single $\mathcal{H}_{E, \delta E}$. This, in turn, will be spanned by entangled states representing a partially evaporated Black Hole, where a fraction $E'$ of the total energy has evaporated into radiation. In other words, $\mathcal{H}_{E, \delta E}$ can be decomposed in terms of factorized Hilbert sub-spaces as
\begin{equation}
    \mathcal{H}_{E, \delta E}=\bigoplus_{E'} \mathcal{H}_{BH, E-E'} \otimes \mathcal{H}_{rad, E'} \, ,
    \label{eq:HilbertDecomposition}
\end{equation}
with the sum over $E'$ to be intended as a continuous integral. Note that the factorization is performed in the gravitational theory, and it corresponds to a factorization in the target space in the language of \cite{Gautam2022}. In this case, emitted $D_0$ branes are completely decoupled from the Black Hole, due to the presence of flat directions, as described in section \ref{sec:BHeva}. This ensures that the factorization in the gravitational theory holds.

We consider a system of $N$ $D_0$-branes which at $t=0$ form a pure Black Hole state (i.e. radiation is in the vacuum state). Due to the finite probability for the Black Hole to radiate a $D_0$-brane, radiation states can be excited. As a consequence, the wave function of the system at time $t$ will be in a superposition, weighted by the probabilities of emission of a given number of $D_0$-branes from an initial number of Black Hole $D_0$-branes. The maximum number of emissions is $N-1$ since the branes can be emitted until only one is left. Then the superposition coefficients are defined as the probabilities $p_0(t),p_1(t),p_2(t),p_3(t),...p_{N-1}(t)$ of $0, 1, 2, 3....N-1$ emissions respectively. These probabilities will be explicitly calculated in the following.

To construct the density matrix, a definition of the Black Hole and radiation states that span the decomposition (\ref{eq:HilbertDecomposition}) is required. Black Hole states are characterized by the number of $D_0$-branes, the invariant energy (mass) in the Black Hole and a recoil momentum that compensate the total momentum of emitted branes. We define these states to satisfy the orthonormality condition:
\begin{equation}
    \prescript{}{BH}{\braket{N,E, \Vec{p}\,|\Tilde{N},\Tilde{E}, \Tilde{\Vec{p}}\,}}_{BH}=\delta_{N,\Tilde{N}}\delta(E-\Tilde{E})\delta(\Vec{p}-\Tilde{\Vec{p}}\,)
\end{equation}
where $\left|N,E, \Vec{p}\,\right>_{BH}$ is a Black Hole stat with $N$ branes, energy $E$ and recoil momentum $\Vec{p}$. 
Using this definition and the decomposition (\ref{eq:HilbertDecomposition}), we can take partial trace over the Black Hole states of any operator $A$ via
\begin{equation}
    A_{rad}=\operatorname{Tr}_{BH}(A)
    =\sum_{i=0}^N\int_0^Edx\int_{\mathbb{R}^9}d\Vec{p}\,\prescript{}{BH}{\left<i,x, \Vec{p}\,\right|}A\left|i,x, \Vec{p}\,\right>_{BH}
\end{equation}
where $E$ is the energy of the black hole at $t=0$, i.e. the total energy of the system, which is conserved in time. This will be necessary in calculating the radiation density matrix.
 
Similarly, we characterize radiation states in (\ref{eq:HilbertDecomposition}) with the number of $D_0$-branes and the nine-momentum of each brane. For example: $\left|m;\Vec{p}^{\,(1)}\, E^{(1)},\Vec{p}^{\,(2)}\,E^{(2)},....\Vec{p}^{\,(m-1)}\,E^{(m-1)},\Vec{p}^{\,(m)}\,E^{(m)}\right>_{rad}$ is a $m$ $D_0$-brane radiation state, which must be intended as a symmetrized product state of individual brane states with fixed momentum. We wrote down the energy dependency explicitly for later convenience, but this is in fact fixed by the dispersion relation $E^{(i)}=\sqrt{m^2_{D0}+\Vec{p}^{\,(i)\, 2}}$. These states are also defined to satisfy the orthonormality condition:
\begin{multline*}
    \prescript{}{rad}{\left<l;\Vec{p}^{\,(1)}\, E^{(1)},...\Vec{p}^{\,(l)}\,E^{(l)}|m;\Bar{\Vec{p}}^{\,(1)}\, \Bar{E}^{(1)},...\Bar{\Vec{p}}^{\,(m)}\,\Bar{E}^{(m)}\right>_{rad}}\\
    =\frac{1}{m!}\delta_{lm} \sum_{\sigma \in \mathcal{S}_m}\prod_{i=0}^l \delta\left(\Vec{p}^{\,(\sigma(i))}-\Bar{\Vec{p}}^{\,(\sigma(i))}\right)\, ,
\end{multline*}
where the sum runs over permutations of the radiation state entries. In the following, we will leverage this Hilbert space construction to calculate the density matrix of the system. This will lead us to the computation of the entanglement entropy.
\section{Page Curve for the BFSS Black Hole}
\label{sec:PageCurve}
The system we want to analyze consists of a BFSS Black Hole, i.e. a bound state of $D_0$-branes, which radiates by emission of individual branes. At $t=0$ the system is in the initial state, namely a Black Hole in a pure state and radiation in the vacuum state
\begin{equation}
\left|\psi(0)\right>=\frac{1}{\sqrt{\delta E}\sqrt{Vol(K)}}\int_{E-\delta E}^Ed\Tilde{E}\int_{K}d\Tilde{\Vec{p}}\,\left|N,\Tilde{E}, \Tilde{\Vec{p}}\right>_{BH} \left|0\right>_{rad} \, .
\label{eq:InitialState}
\end{equation}
The Black Hole pure state is a superposition of energy eigenstates within a narrow micro-canonical window. Recoil momentum is also smeared within a compact subset $K \subset \mathbb{R}^9$ centered around $\Vec{p}=0$ and with a small radius $\delta p$. Time evolution for the system is unitary, therefore the overall state will remain pure. In principle, such a time evolution is the result of the action of the BFSS Hamiltonian, namely the Legendre transform of (\ref{eq:BFSS}), in the matrix theory, which results in the time evolution of the initial state (\ref{eq:InitialState}). However, a full description of the Black Hole state purely in terms of Matrix Theory quantities is not known, hence the action of the full Hamiltonian on this state cannot be determined exactly. Therefore, we have to resort to an effective description. In particular, we describe the state at later times $t$ as the superposition of partially evaporated Black Hole plus radiation states, the weighting coefficients being by the time-dependent probabilities of emission of branes, $p_i(t)$. It follows that the state of the system at time $t$ takes the form
\begin{multline}
    \left|\psi(t)\right>=\sum_{m=0}^{N-1}\frac{p_m(t)^{1/2}a_m(E-\delta E)}{\sqrt{\delta E}\sqrt{Vol(K)}} \int_{E-\delta E}^Ed\Tilde{E}\int_{K}d\Tilde{\Vec{p}} \\
    \int_0^{E-\delta E} \left[d\Vec{p}
    \,\right]^{(m)}
    \left|N-m,\Tilde{E}-\sum_{i=1}^mE^{(i)}-E^K_{BH}, \Tilde{\Vec{p}}-\sum_{i=1}^m\Vec{p}^{\,(i)}\right>_{BH}\\
    \otimes \, \left|m;\Vec{p}^{\,(1)}\, E^{(1)},\Vec{p}^{\,(2)}\,E^{(2)},....\Vec{p}^{\,(m-1)}\,E^{(m-1)},\Vec{p}^{\,(m)}\,E^{(m)}\right>_{rad}
    \label{eq:Statet}
\end{multline}
where $E^K_{BH}$ is the Black Hole recoil kinetic energy and we defined
\begin{equation}
    \int_0^x\left[d\Vec{p}\,\right]^{(m)}
    =\left\{\begin{matrix}
1 & \text{for}\ m=0\\[5pt]
\prod_{i=1}^m\int d\Omega_9^{(i)}\int_{m_{D_0}}^{\infty} dE^{(i)}\, E^{(i)}(E^{(i)\, 2}-m^2_{D_0})^{7/2}\\\times\, \theta(x-\sum_{i=1}^mE^{(i)}-E^K_{BH}) & \text{for}\ m\neq 0
\end{matrix}\right.
\end{equation}
where $d\Omega_9^{(i)}$ is the nine-dimensional solid angle for each brane and the $\theta$ function ensures that the energy extracted from the Black Hole by radiation does not exceed the initial Black Hole mass. In particular, it ensures that the integral has support on a compact domain, and therefore converges. The normalization constants $a_m(E-\delta E)$ are fixed by requiring
\begin{equation}
   a_m(x)^2\int_0^x\left[d\Vec{p}\,\right]^{(m)}=1  \, .
\end{equation}
Note that
\begin{multline}
    \left<\psi(t)|\psi(t)\right>
    =\sum_{m=0}^{N-1}\frac{p_m(t)a_m(E-\delta E)^2}{\delta E \,Vol(K)}\int_{E-\delta E}^Ed\Tilde{E}d\Tilde{E'}\int_{K}d\Tilde{\Vec{p}}\,'\int_{K}d\Tilde{\Vec{p}} 
   \times \int_0^{E-\delta E} \left[d\Vec{p}\,\right]^{(m)}\left[d\Bar{\Vec{p}}\,\right]^{(m)}\\
    \times \frac{1}{m!} \sum_{\sigma \in \mathcal{S}_m}\prod_{m=0}^l \delta\left(\Vec{p}^{\,(\sigma(m))}-\Bar{\Vec{p}}^{\,(\sigma(m))}\right)
    \times \delta(\Tilde{E}-\Tilde{E'}-\sum_{i=1}^m E^{(i)}+\sum_{i=1}^m\Bar{E}^{(i)}-E^{K}_{BH}+\Bar{E}^{K}_{BH})\\
    \times \delta(\Tilde{\Vec{p}}-\Tilde{\Vec{p}}\,'-\sum_{i=0}^m \Vec{p}^{\,(i)}+\sum_{i=0}^m \Bar{\Vec{p}}^{\,(i)}) \\
    =\sum_{m=0}^{N-1}\frac{p_m(t)a_m(E-\delta E)^2}{\delta E\, Vol(K)}
    \times \int_{E-\delta E}^Ed\Tilde{E}\int_{K}d\Tilde{\Vec{p}}\int_0^{E-\delta E} \left[d\Vec{p}\,\right]^{(m)}=1
    \label{eq:Norm}
\end{multline}
meaning that the state is properly normalized at all times. 

Since the system is in a pure state given by (\ref{eq:Statet}), the density matrix of the system reads
\begin{equation}
\rho(t)=\left|\psi(t)\right>\left<\psi(t)\right| 
\label{eq:density}
\end{equation}
and the normalization property (\ref{eq:Norm}) implies $\operatorname{Tr}\rho(t)=1$.
To compute the entanglement entropy of the radiation, we need to consider the partial trace of the density matrix (\ref{eq:density}) with respect to Black Hole states, thus obtaining the reduced density matrix:
\begin{multline*}
    \rho_{rad}(t)=\operatorname{Tr}_{BH}\rho(t)
    =\sum_{m=0}^{N-1}\int_0^Edx\int_{\mathbb{R}^9}d\Vec{p}\,\,_{BH}\left<m,x,\Vec{p}\,\right|\rho(t)\left|m,x, \Vec{p}\,\right>_{BH}\\
=p_0(t)\,\,_{rad}\left|0\right>\left<0\right|_{rad}
+\sum_{m=1}^{N-1} \int_0^{E}dx\int_{\mathbb{R}^9}d\Vec{p}\,\,\frac{p_m(t)}{\delta E \,Vol(K)}a_m(E-\delta E)^2\\
\times \int_{E-\delta E}^Ed\Tilde{E}d\Tilde{E'}\int_{K}d\Tilde{\Vec{p}}\,'\int_{K}d\Tilde{\Vec{p}}\int_0^{E-\delta E} \left[d\Vec{p}\,\right]^{(m)}\left[d\Bar{\Vec{p}}\,\right]^{(m)}\\
\times \delta(x-\Tilde{E}+\sum_{i=1}^m E^{(i)}+E^K_{BH})
 \delta(x-\Tilde{E'}+\sum_{i=1}^m \Bar{E}^{(i)}+\Bar{E}^K_{BH})\\
 \times \delta(\Vec{p}-\Tilde{\Vec{p}}+\sum_{i=1}^m\Vec{p}^{\,(i)})\delta(\Vec{p}-\Tilde{\Vec{p}}\,'+\sum_{i=1}^m\Bar{\Vec{p}}^{\,(i)})\\
 _{rad}\left|m;\Vec{p}^{\,(1)}\,E^{(1)},...\right>\left<m;\Bar{\Vec{p}}^{\,(1)}\,\Bar{E}^{(1)},...\right|_{rad} 
 \end{multline*}
 which can be simplified by solving the delta function integrals, thus obtaining
 \begin{multline*}
\rho_{rad}(t)=\sum_{m=0}^{N-1} p_m(t)a_m(E-\delta E)^2\int_0^{E-\delta E} \left[d\Vec{p}\,\right]^{(m)}\left[d\Bar{\Vec{p}}\,\right]^{(m)} \\
_{rad}\left|m;\Vec{p}^{\,(1)}\,E^{(1)},...\right>\left<m;\Bar{\Vec{p}}^{\,(1)}\,\Bar{E}^{(1)},...\right|_{rad} \, .
\end{multline*}
We note that we can define superposition states for radiation with fixed particle number
\begin{equation}
    \left|i\right>_{rad}:=
    \int_0^{E-\delta E} \left[d\Vec{p}\,\right]^{(i)}a_i(E-\delta E)\left|i;\Vec{p}^{\,(1)}\,E^{(1)},....,\Vec{p}^{\,(m)}\,E^{(i)}\right>_{rad} \, ,
\end{equation}
which satisfy the orthonormality condition
\begin{equation}
    \left._{rad}\left<i|j\right>_{rad}\right.=\delta_{i,j} \, .
\end{equation}
With this choice of basis, the density matrix takes a particularly simple form 
\begin{equation}
    \rho_{rad}(t)=\sum_{i=0}^{N-1} p_i(t)\,\,_{rad}\left|i\right>\left<i\right|_{rad}\, .
\end{equation}
As a consequence, computing von Neumann entanglement entropy is straightforward, and it yields
\begin{equation} 
    S(t)=-\operatorname{Tr}(\rho_{rad}(t)\ln {\rho_{rad}(t)})=-\sum_{i=0}^{N-1} p_i(t)\operatorname{ln}\left(p_i(t)\right) \, ,
    \label{eq:entropy}
\end{equation}
and the purity of the radiation state is given by
\begin{equation} 
    P(t)=-\operatorname{Tr}(\rho^2_{rad}(t))=-\sum_{i=0}^{N-1} p_i^2(t) \, .
    \label{eq:purity}
\end{equation}
All in all, we have obtained closed expressions for the entanglement entropy and the purity of the radiation state in terms of the probabilities $p_i(t)$ of emitting $D_0$-branes by Hawking radiation, namely the probability for flat directions of the Matrix Theory potential to open up. In the following, we will obtain an explicit expression for those probabilities. This will enable us to fully characterize the profile of the entanglement entropy and purity as functions of time.

We consider the emission of branes by Hawking radiation (through the mechanism outlined in section \ref{sec:BHeva} ) as a sequence of radiative processes. Defining the mean emission time of a brane from a  composed of $N-i$ $D_0$-branes as $t_i$, the probability of no emission $p_0$ is given by
\begin{equation}
    p_0(t)=e^{-t/t_0} \, .
\end{equation}
We can use this to compute the probability of one emission $p_1$. This is given by
\begin{equation}
    p_1(t)=A_1\int_0^t dT \,p_{0}(T)e^{(T-t)/t_1} \, ,
\end{equation}
where $A_1$ is a $t$-independent factor such that the probability of one emission from $N$ branes in infinitesimal time $\delta t$ is $A_1 \delta t$. This can be generalized to recursively determine the probabilities $p_i$ of $i$ emissions as
\begin{equation}
    p_i(t)=A_i\int_0^t dT \,p_{i-1}(T)e^{(T-t)/t_i} \, ,
    \label{eq:ProbRecursion}
\end{equation}
where $A_i$ are $t$-independent factors such that the probability of one emission from $N-i+1$ branes in infinitesimal time $\delta t$ is $A_i \delta t$. These factors are found by imposing the consistency condition on probabilities
\begin{equation}
    \sum_{i=0}^{N-1}p_i(t)=1 \, .
    \label{eq:consistency}
\end{equation}
However, we cannot have more than $N-1$ emissions, since after emitting $N-1$ $D_0$-branes, there is only one $D_0$-brane left, meaning that the Black Hole has already evaporated into $N$ non-interacting branes. It follows that $p_N(t)=0$, hence $t_{N-1}\rightarrow\infty$. As a consequence, (\ref{eq:consistency}) is translated into
\begin{equation} \label{eq:NormCondition}
    A_{N-1}\int_0^tdTp_{N-2}(T)=1-\sum_{i=0}^{N-2}p_i(t) \, .
\end{equation}
The condition (\ref{eq:NormCondition}) can be solved recursively to obtain the normalization factor: 
\begin{equation} \label{eq:proof}
    A_i=\frac{1}{t_{i-1}} \, ,
\end{equation}
and the steps leading to this equation are carried out explicitly in Appendix \ref{sec:Norm}.

\begin{figure}[htbp]
\centering
\includegraphics[width=0.5\textwidth]{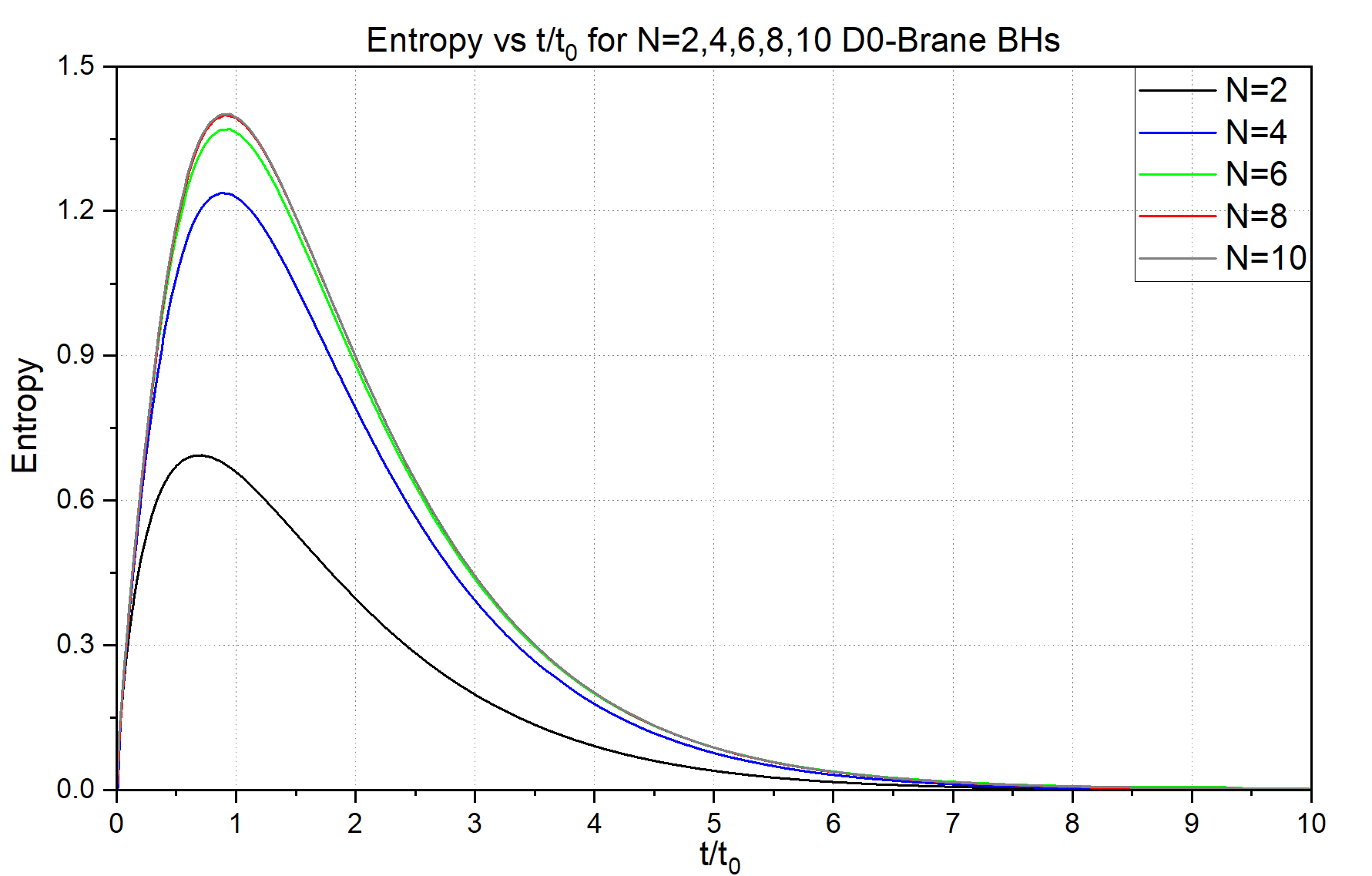}
\caption{\small{Entanglement entropy for the radiation as a function of time.}}
\label{fig:Entropy_Time1}
\end{figure}

\begin{figure}[htbp]
\centering
\includegraphics[width=0.52\textwidth]{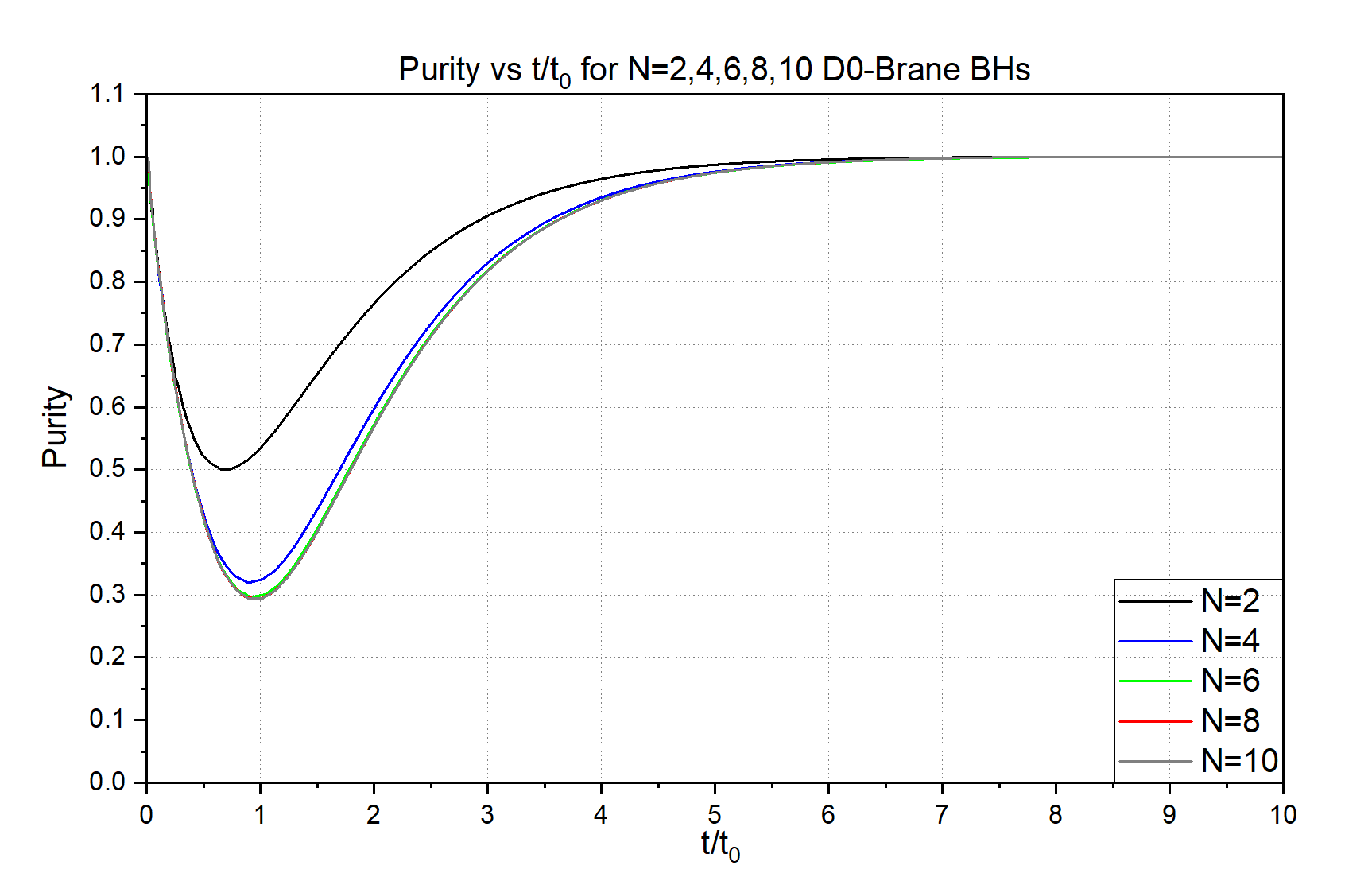}
\caption{\small{Purity of the radiation density matrix as a function of time.}}
\label{fig:Purity_Time1}
\end{figure}

Having derived the emission probabilities, we now have full access to the entropy and purity as functions of time, equations (\ref{eq:entropy}) and (\ref{eq:purity}) respectively, which we can plot for different values of $N$. Assuming the timescale for the emission of a brane from a partially evaporated Black Hole to be 
\begin{equation} \label{eq:EmissionTime1}
    t_i \approx e^{N-i}
\end{equation}
as argued in \cite{Berkowitz_2016}, the results for the plots are shown in Figures \ref{fig:Entropy_Time1} and \ref{fig:Purity_Time1}.
From these graphs, we come to the following conclusions:
\begin{itemize}
    \item Unitarity of the Black Hole evaporation process is preserved since the entropy tends to 0 as $t\rightarrow\infty$, i.e. after the Black Hole has fully evaporated;
    \item The information that was lost after the creation of the Black Hole is recovered after complete evaporation of the Black Hole in the form of Hawking radiation in a pure state. This is the behaviour displayed in Figure \ref{fig:Purity_Time1}, where we see that the purity of the radiation goes back to $0$ after evaporation;
    \item The peak of the radiation entropy curve occurs at the Page time. This is defined as the time for which the Black Hole has radiated half of its mass. We note that $t_P \sim O(t_0)$.
    \end{itemize}
Note that the average number of emitted branes is 
\begin{equation}
    \braket{n(t)}=\sum_{k=0}^{N-1} k p_k(t) \, .
\end{equation}
Moreover, by differentiating (\ref{eq:ProbRecursion}), we get 
\begin{equation}
    \frac{dp_i(x)}{dx}+\frac{p_i(x)}{t_i}=\frac{p_{i-1}(x)}{t_{i-1}} 
\end{equation}
and, as a consequence 
\begin{equation}
    \frac{d}{dt} \braket{n(t)}=\sum_{k=0}^{N-1} \frac{1}{t_k} p_k(t)=\braket{\,\frac{1}{t_n}} \, .
\end{equation}
Considering that the number of branes in the Black Hole is given by $n_{BH}=N-n$, it follows that
\begin{equation}
    \frac{d}{dt} \braket{n_{BH}(t)}=-\braket{\,\frac{1}{t_{N-n_{BH}}}} \, .
    \label{eq:QuantumRad}
\end{equation}
If we ignore quantum fluctuations and restrict ourselves to the classical picture presented in \cite{Berkowitz_2016}, equation (\ref{eq:QuantumRad}) reads
\begin{equation}
    \frac{d \Bar{n}}{d t}=-e^{-\Bar{n}} \, ,
\end{equation}
where $\Bar{n}=\braket{n_{BH}}$, which in turn implies 
\begin{equation}
    \int_N^{\Bar{n}} d n e^n=\int_0^t-d t \Longrightarrow t=e^N-e^{\Bar{n}} \, .
\end{equation}
If $t_{1 / 2}$ is the half-life of the Black Hole,
\begin{equation}
    \frac{t_{1 / 2}}{t_0}=\frac{e^N-e^{N / 2}}{e^N}=1-e^{-N / 2} \, .
\end{equation}
Hence, as $N\rightarrow \infty$,
\begin{equation}
    \frac{t_{1/2}}{t_0} \longrightarrow 1 \,
\end{equation}
namely, the Black Hole half-life is equal to $t_0$ in the large $N$ limit. The analogous quantity to the half-life in our quantum picture is the Page time and we see that the classical approximation captures the $t_P\sim O(t_0)$ behaviour that we observed in our analysis. This can be seen as a consistency check of the classical limit in \cite{Berkowitz_2016}, since qualitative features such as the scaling of the Page time with $t_0$ do not change dramatically in the quantum picture. Of course, this is not the end of the story. Describing the Black Hole state and its radiation purely in terms of Matrix Theory quantities would provide us with a full understanding of the problem and, therefore, remains an interesting route to investigate.
\section{A Quantitative Estimate of the Emission Time}
\label{sec:Eigendensity}
The approximate form of the mean decay time given in \cite{Berkowitz_2016}, namely $t \approx e^{N}$ where $N$ is the number of $D_0$-branes in the Black Hole, is inferred from the relation
\begin{equation}
    t\approx e^{\Delta S}
\end{equation}
where $\Delta S$ is the entropy difference of the (BH + no radiation) and the (BH + 1 brane) configurations. A more quantitative estimate for the entropy difference can be found by considering the distribution of the largest eigenvalue of the matrix $R=\sqrt{X_IX^I}$, defined in terms of the BFSS position matrices $X^I$. This was also computed in \cite{Berkowitz_2016}, where they show that it takes the form
\begin{equation}
\rho(r) \sim(\lambda T)^{-1 / 4} \cdot\left[\frac{r}{(\lambda T)^{1 / 4}}\right]^{-8(2 N-3)} \, ,
\label{eq:RadiusDistr}
\end{equation}
where $\lambda=Ng^2_{YM}$ is the 't Hooft coupling and we are considering the theory at finite temperature $T$. Here $\rho(r)$ can be thought of as the distribution of the radial coordinate of the $D_0$-brane furthest away from the Black Hole. The distribution (\ref{eq:RadiusDistr}) represents the leading contribution in the large $N$ limit and might be affected by $O(\frac{1}{N})$ corrections at finite $N$. Since, as described in section \ref{sec:QuantSystem}, we are restricting to a micro-canonical window, we can write the entropy of a state as $S=\operatorname{log}(\Omega)$ where $\Omega$ is the phase space volume corresponding to the $D_0$-brane configuration of the state. In particular, the entropy of the configuration in which no brane has been emitted, which we denote by $S_<$, is computed from the phase space integral of the distribution (\ref{eq:RadiusDistr}). The limits of integration must be chosen to ensure that the $D_0$-brane whose radial coordinate is distributed according to (\ref{eq:RadiusDistr}) is the furthest away from the singularity while still interacting with the Black Hole. Hence, we translate the statement of $r$ being the largest eigenvalue of the radial coordinate matrix as the lower bound of integration in (\ref{eq:entropyNoEm}). Then, $r_0$ is the average radius of the region occupied by the bound state of the remaining $N-1$ branes. $R$ is the radius at which the $D_0$-brane is considered to decouple. It follows that, in this case,
\begin{equation}
    e^{S_<}=C\int_{r_0}^Rdr\rho(r) 
    \label{eq:entropyNoEm}
\end{equation}
where the factor $C$ accounts for the phase space in the remaining $8$ directions and is, therefore, independent of the radial coordinate.
Similarly, the entropy of the configuration in which a $D_0$-brane has decoupled from the Black Hole is determined through the equation
\begin{equation}
    e^{S_>}=C\int_{R}^\infty dr\rho(r)
\end{equation}
where $S_>$ is the entropy for such a configuration. Hence, the suppression factor is given by 
\begin{equation}
    e^{\Delta S}=\frac{e^{S_<}}{e^{S_>}}=\frac{\int_{r_0}^Rx^{-8(2 N-3)}dx}{\int_{R}^\infty x^{-8(2 N-3)}dx} =e^{(8(2N-3)-1)\operatorname{log}(R/r_0)}-1 \, .
\end{equation}
The above corresponds to the mean emission time of a brane from a bound state of $N$ $D_0$-branes. Indeed this corresponds to a more quantitative version of (\ref{eq:EmissionTime1}), which yields
\begin{equation} \label{eq:EmissionTime2}
    t_i=e^{(8(2(N-i)-3)-1)\operatorname{log}(R/r_0)}-1
\end{equation}
as a mean emission time.

\begin{figure}[htbp]
\centering
\includegraphics[width=0.53\textwidth]{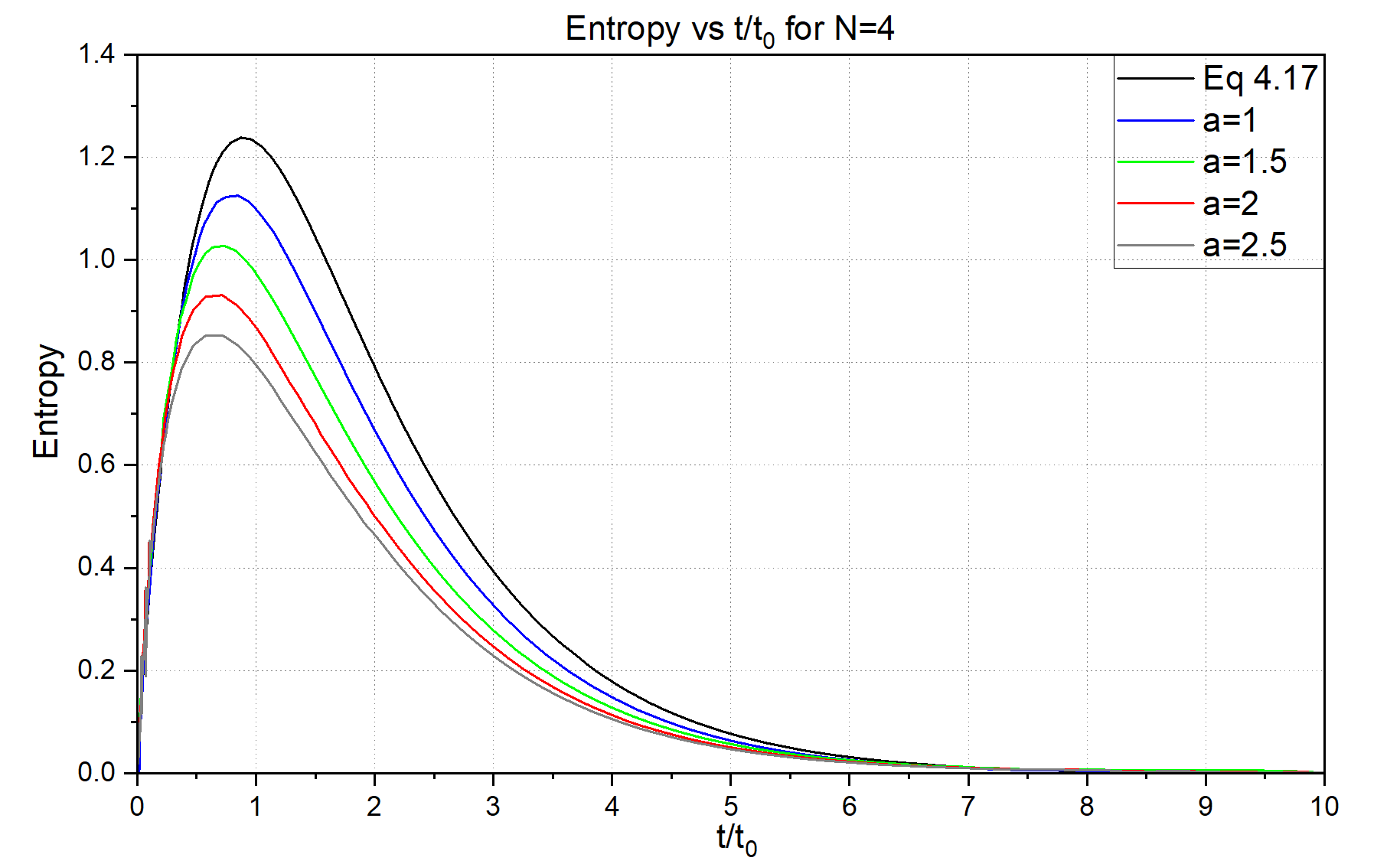}
\caption{\small{Refined entropy vs time graph for a 4 D0-brane BH. Equation (\ref{eq:EmissionTime2}) is the mean emission time with $a=1, 1.5, 2, 2.5$ and is compared with the curve when considering equation (\ref{eq:EmissionTime1}) as the mean emission time.}}
\label{fig:Entropy_Time2}
\end{figure}

\begin{figure}[htbp]
\centering
\includegraphics[width=0.53\textwidth]{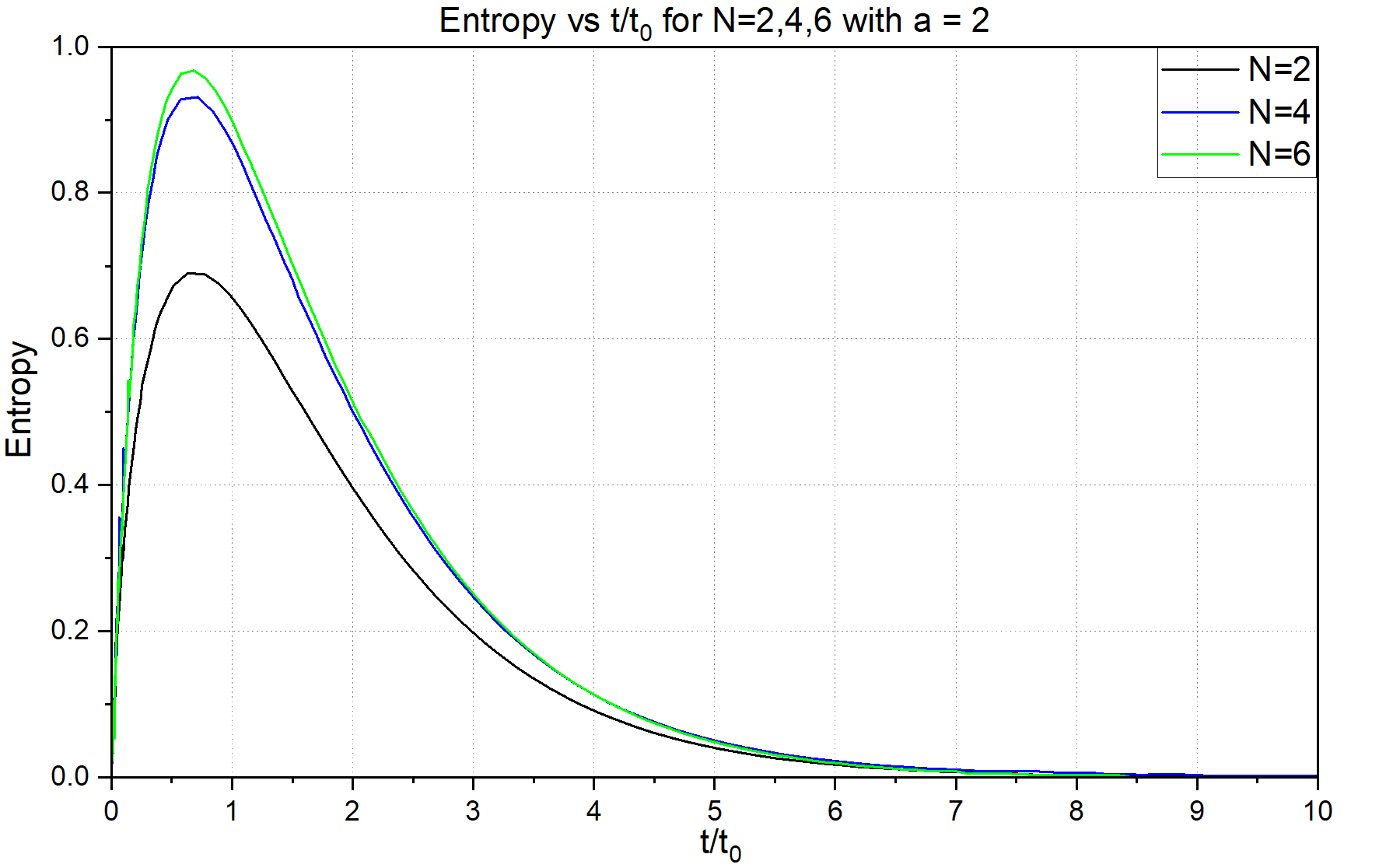}
\caption{\small{Refined entropy vs time graph. The plot shows entropy graphs obtained by considering equation (\ref{eq:EmissionTime2}) as the mean emission time equation with $a=2$ and $N=2, 4, 6$.}}
\label{fig:Entropy_Time3}
\end{figure}

Having obtained a more quantitative estimate for the emission times, equation (\ref{eq:EmissionTime2}), we now repeat the analysis carried out in section \ref{sec:PageCurve} and compute a refined version of the Page curve. To this end, it is useful to define the new parameter
\begin{equation}
    a=16\ln(R/r_0)\, .
    \label{eq:aparam}
\end{equation} 
We plug equation (\ref{eq:EmissionTime2}) in (\ref{eq:entropy}) and we look at plots of the entropy as a function of time. Those are shown in Figures \ref{fig:Entropy_Time2} and \ref{fig:Entropy_Time3} for different values of the parameters.
From the graphs, we infer the following:
\begin{itemize}
    \item As before, the entropy comes back to 0 after the evaporation of the Black Hole for all cases, showing the expected Page curves. Again the BFSS Black Hole represented with our model does not feature any information loss;
    \item The Page time reduces with an increase in $a$. This matches with the conclusions one can draw from the half-life of the Black Hole in the classical limit, given below;
 \item As $a$ increases, the entropy reduces at all times. 
\end{itemize}
In this case, the classical limit of equation (\ref{eq:QuantumRad}) reads
\begin{equation}
    \frac{d\Bar{n}}{dt}=\frac{-1}{e^{a\Bar{n}-b}-1}
\end{equation}
where $a=16\ln(R/r_0)$ and $b=\frac{25a}{16}$.
Applying the appropriate boundary conditions, we get
\begin{equation}
    t=\frac{1}{a}\left(e^{aN-b}-e^{a\Bar{n}-b}\right)-N+\Bar{n} \, ,
\end{equation} 
which implies for the Black Hole half-life
\begin{equation}
    t_{1/2}=\frac{1}{a}\left(e^{aN-b}-e^{a\frac{N}{2}-b}\right)-\frac{N}{2} \, .
\end{equation}
Hence, the ratio $\frac{t_{1/2}}{t_0}$ is given by
\begin{equation}
    \frac{t_{1/2}}{t_0}=\frac{e^{aN-b}-e^{a\frac{N}{2}-b}-\frac{aN}{2}}{a(e^{aN-b}-1)} \, .
\end{equation}
In particular, we note that as $a$ increases, this ratio decreases. Moreover, for large values of $N$ 
\begin{equation}
    \frac{t_{1/2}}{t_0} \longrightarrow \frac{1}{a} \, ,
\end{equation}
which is consistent with what we observe for the Page time in Figure \ref{fig:Entropy_Time2}. Once again, the classical approximation captures the qualitative features that we observe in the quantum system.

All in all, we obtained a profile for the entanglement entropy as a function of time which depends on the features of the Black Hole state through the parameter $a$.
   \section{Conclusion and Outlook}
   \label{sec:Conclusion}
   In this paper, we have considered the Black 0-Brane solution to supergravity, which is dual to BFSS Matrix theory, and we have computed the entanglement entropy of Hawking radiation. The presence of a Gauge theory dual allows for a microscopical description of the Black Hole evaporation mechanism. In particular, the Black Hole state is dual to a bound state of branes clumped around the origin and held together by open strings. Evaporation is then caused by the emission of branes which run astray from the clump. Flat directions in the potential, together with the chaotic nature of the theory, make this possible, as the interaction between branes is turned off at a large distance due to a supersymmetric cancellation. The process of evaporation is, therefore, fully resolved in terms of unitary evolution in the dual Gauge theory description and no information should go missing after the Black Hole is completely evaporated. We checked this statement by computing the entanglement entropy resulting from quantizing the radiation process and showed the expected Page curve is indeed reproduced. The computation is based on assuming the Hilbert space to factorize into partially evaporated Black Hole plus radiation configurations. We then considered a more quantitative description of the rate of emission of branes based on the distribution of the largest eigenvalue of the radial coordinate matrix $R=\sqrt{X^I X_I}$ proposed in \cite{Berkowitz_2016}. Hence we found a more refined version of the entanglement entropy as a function of time, which also reproduces the desired Page curve. As such, this computation confirms the fact that the mechanism put forth in \cite{Berkowitz_2016, Hanada_2016} for $D_0$-branes emission indeed represents Hawking radiation in the gravitational dual description.

   Interestingly, the quantitative estimate for the entanglement entropy presented in section \ref{sec:Eigendensity} depends on the parameter $a$ in (\ref{eq:aparam}), roughly characterizing the size of the Black Hole. To make a connection with the geometric interpretation presented in \cite{Berkowitz_2016}, $r_0$, which is the size of the $N$ $D_0$-brane clump, can be thought of as the size of the resolved singularity, whereas the radius $R$, representing the distance at which branes are considered to decouple and be emitted, might characterize the Black Hole horizon. It would be interesting to further probe this connection.
   
   The mechanism for quantum evaporation we considered is the quantized version of a classical process. As such, our results stand as an important consistency check for the interpretation of a bound state of $D_0$-branes as an evaporating Black Hole. However, a fully quantum description of Black Hole evaporation, as well as of the Black Hole state itself, is not known. An interesting route to explore would be to try to fully determine the Black Hole state in terms of Matrix Theory quantities. This would certainly improve our understanding of the entanglement entropy to include $O(\frac{1}{N})$ corrections and would shed light on the geometric interpretation of the model.

  \section*{Acknowledgement}
The authors would like to thank John Wheater for useful discussions and comments on the draft of the present work. We thank Gabriel Wong for comments on the draft of the present work. We thank Ludovic Fraser-Taliente for useful discussions and comments on version one of the pre-print. This work has been partially funded by STFC, studentship grant ST/X508664/1. For the purpose of open access, the author has applied a CC
BY public copyright licence to any Author Accepted Manuscript
(AAM) version arising from this submission.
\appendix
\section{Normalization factor for Probabilities}
\label{sec:Norm}
In this section, we provide a proof by induction of equation (\ref{eq:proof}), namely that the correct normalization factor for probabilities is $A_i=\frac{1}{t_{i-1}}$, $\forall i=1,...,d$. We proceed by induction on $d=N-1$. For $d=1$, we simply integrate equation (\ref{eq:NormCondition})
\begin{equation}
    p_1(t)=A_1\int_0^t p_0(t)=1-p(0)
\end{equation}
Using $p_0(t)=e^{-t/t_0}$, we get $A_1=1/t_0$. Now, assume that equation (\ref{eq:proof}) holds for $i=1,...,d$. We have to prove that $A_{d+1}=1/t_d$. For this, consider equation (\ref{eq:ProbRecursion}), which we rewrite here for convenience, under the induction hypothesis
\begin{equation}
    p_i(x)=\frac{1}{t_{i-1}}\int_0^xdt\ p_{i-1}(t)e^{(t-x)/t_i} \, .
\end{equation}
Differentiating the above yields
\begin{equation} \label{eq:differential}
    \frac{dp_i(x)}{dx}+\frac{p_i(x)}{t_i}=\frac{p_{i-1}(x)}{t_{i-1}} \, .
\end{equation}
In order to find $A_{d+1}$, we consider equation (\ref{eq:NormCondition}), which now reads
\begin{equation}
   A_{d+1} \int_0^xdt\ p_{d}(t)=1-\sum_{i=0}^{d}p_i(x) \, .
\end{equation}
We differentiate the above with respect to $x$ and we obtain
\begin{equation}
    A_{d+1} \,p_{d}(x)=-\sum_{i=0}^{d}\frac{dp_i(x)}{dx}= \sum_{i=0}^d\frac{p_{i}(x)}{t_{i}}-\sum_{i=1}^d\frac{p_{i-1}(x)}{t_{i-1}}=\frac{p_{d}(x)}{t_{d}} \, ,
\end{equation}
which implies
\begin{equation}
    A_{d+1}=\frac{1}{t_d}
\end{equation}
as required.

\bibliographystyle{unsrturl}
\bibliography{references.bib}

\end{document}